\documentclass[twocolumn,showpacs,preprintnumbers,amsmath,amssymb]{revtex4}
\usepackage{epsfig}
\usepackage{graphicx}% Include figure files
\usepackage{dcolumn}
\usepackage{bm}
\newcommand{\be}{\begin{eqnarray}}
\newcommand{\ee}{\end{eqnarray}}

\newcommand{\ave}[1]{\left\langle #1 \right\rangle}

\begin{document}\hbadness=10000
%\topmargin=-.7cm\oddsidemargin = -0.7cm\evensidemargin = -1.7cm
%\draft
%%%%%%%%%%%%%%%%%%%%%%%%%%%%%%%%%%%%%%%%%%%

\title{A statistical model analysis of yields and fluctuations in 200 GeV Au-Au collisions}
\author{Giorgio Torrieri}
\affiliation{
Department of Physics, McGill University, Montreal, QC H3A-2T8, Canada}
\author{Sangyong Jeon}
\affiliation{
Department of Physics, McGill University, Montreal, QC H3A-2T8, Canada}
\affiliation{
RIKEN-BNL Research Center, Upton NY, 11973, USA}
\author{Johann Rafelski}%,
\affiliation{
Department of Physics, University of Arizona, Tucson AZ 85721, USA}
\date{September, 2005}

\begin{abstract}
We show that the simultaneous measurement of yields and fluctuations is capable of falsifying and constraining the statistical hadronization model.   We show how such a measurement can test for chemical non-equilibrium, and distinguish between a high temperature chemically equilibrated freeze-out from a supercooled freeze-out with an over-saturated phase space.
We perform a fit, and show that both yields and fluctuations measured at RHIC 200 GeV can be accounted for within the second scenario, with both the light and strange quark phase space saturated significantly
above detailed balance.
%We show that this model can fit $K/\pi$ fluctuations, pion, proton and hyperon ratios, and resonances measured at the STAR experiment.
We point to the simultaneous fit of the $K/\pi$ fluctuation and the $K^*/K^-$ ratio as evidence that the effect of hadronic re-interactions after freeze-out is small.
\end{abstract}
%%%%%%%%%%%%%%%%%%%%%%%%%%%%%%%%%%%%%%%%%%%%

\maketitle
The statistical
hadronization model (SHM) 
\cite{Fer50,Pom51,Lan53} has been extensively applied to the study of soft particle production in hadronic systems.   When it includes the full spectrum of hadronic
resonances \cite{Hag65}, the SHM can describe quantitatively
the abundances of all light and strange hadrons produced in heavy ion collisions.   

The ability of the SHM to describe not just averages, but event-by-event multiplicity fluctuations
has however not been widely investigated, and its applicability is currently a matter of controversy.
Event-by-event particle fluctuations have elicited 
theoretical~\cite{fluct2,jeonkochratios,fluct4,fluct6}
and experimental  interest~\cite{starfluct2,phefluct}, both as a consistency check
for existing models ~\cite{fluct2,jeonkochratios}, as a test of hadron gas equilibration and the presence of
resonances \cite{fluct2}, and as a way to search  for new physics \cite{fluct4}.

This study investigates weather the SHM can explain both yields and fluctuations with the same parameters.
We illustrate the use of both yields and fluctuations as a probe capable of differentiating between freeze-out scenarios.
We then perform a fit to preliminary 200 GeV RHIC experimental data, and discuss the results in the context of chemical (non)equilibrium within the RHIC system, the freeze-out temperature and
the presence (or absence) of hadronic reinteraction after freeze-out.

The statistical hadronization model assumes that particles are produced according to a probability determined by their phase
space density.   The first and second cumulants of this probability distribution give, respectively, the average value (over all events) of the desired observable, and its event-by-event fluctuation.

Conserved quantities can be treated in several ways, appropriate to different experimental situations:  If the totality of the system is observed, than conserved quantities
can not fluctuate.   If a small fraction equilibrated with the rest of the system ( the ``bath'')  is observed, than conserved quantities will fluctuate, event-by-event.
Rigorous conservation is known as the Microcanonical ensemble, while allowing energy and other conserved quantities to fluctuate between the system and the bath leads, respectively, to the Canonical and Grand Canonical (GC) ensembles.    As shown recently \cite{nogc2}, \textit{all} fluctuations are ensemble-specific even in the thermodynamic limit.

 In this work, we use the GC ensemble, implemented in a publicly-available software package \cite{share} to calculate fluctuations and yields.  We motivate this choice by the fact that RHIC experiments
observe the mid-rapidity slice of the system, comprising roughly 1/8 of the total multiplicity, an appropriate fraction for a GC prescription.  Boost invariance, a good symmetry around mid-rapidity, links this rapidity slice with a sector in configuration space.
If the system observed at RHIC is a nearly ideal fluid, the matter created in this space should be in equilibrium, grand-canonically, with the unobserved regions.
    If freeze-out temperature throughout observed space is approximately constant, the GC ensemble should be able to describe both yields and fluctuations \cite{cleymans,prlfluct}.

The final state yield of particle $i$ is computed by adding the direct
yield and all resonance decay feed-downs.
\be
 \label{resoyield}
\langle N_i\rangle_{\rm total} & = &
\langle N_i\rangle + \sum_{{\rm all}\;j \rightarrow i}
B_{j \rightarrow i}  \langle   N_j\rangle 
 \label{fluctdef}
 \ee
Here $\ave{N_i}$ is the ensemble average of the $i$-th particle (resonance)
multiplicity at chemical freeze-out and
$B_{j\to i}$ is the branching ratio of the $j \rightarrow i$ decay.
The fluctuation $\ave{\Delta X^2}=\ave{X^2}-\ave{X}^2$ also has two parts
\be
\ave{\Delta N_{j\to i}^2}
& = &
B_{j\to i}(1-B_{j\to i})\ave{N_j}
+
B_{j\to i}^2 \ave{\Delta N_j^2}\phantom{.....}
\ee
assuming no primordial correlation between $i$ and $j$.

The fluctuation of a ratio $N_1/N_2$ (a useful experimental observable, since volume fluctuations cancel out event-by-event)
can be computed from the fluctuation of the denominator and the numerator \cite{jeonkochratios}
($\sigma_{X}^2=\ave{\Delta X^2}/\ave{X}$):
\be
\label{fluctratio}
\sigma_{N_1/N_2}^2
&=&\frac{1}{\ave{N_{2}}}\left( D_{11}+D_{22}-2D_{12}\right)\\
D_{11} & = & {\ave{N_2}\over \ave{N_1}}\, F_1  \phantom{A},\phantom{A}D_{22}
 =  F_2
\\
D_{12}
& = &
\sum_{j \rightarrow 1,2} B_{j \rightarrow 1,2} {{\ave{N_j}}\over {\ave{N_1}}}
\label{eq:d_+-}\\
 F_i &=& \left(
\frac{ \ave{\Delta N_{i}^2}_{\rm direct}}{\ave{N_i}}
+\sum_{j \rightarrow i}\ave{N_{j\rightarrow i}^{2}} \frac{\ave{N_{j \rightarrow i}}}{\ave{N_i}}\right) \phantom{......}
\ee
Note the appearance of a negative correlation term $D_{12}$, which arises due to the correlation between $N_1$ and $N_2$ stemming from a common
resonance feed-down ($\Delta \rightarrow p \pi$ will be a source of correlation between $N_{p}$ and $N_{\pi}$).

The average yields and fluctuations of hadrons themselves can be calculated by a textbook
method given a grand canonical
description of the phase space;   For a hadron with an energy
$E_{p} = \sqrt{p^2+m^2}$, the average number, and event by event fluctuation will be given by
\be
\label{yield_formula}
\langle N_i\rangle
  &=&  gV\int {d^3p\over (2\pi)^3}\, {1\over  \lambda_i^{-1} e^{E_p\beta}\pm 1}\\\langle\Delta N_i^2\rangle & = & \lambda_i{\partial \over
                 \partial \lambda_i} \langle N_i\rangle
\ee
where the upper sign is for Fermions and the lower sign is for  Bosons.
Here $\lambda_i$ is the fugacity of particle $i$, see Eq.\,(\ref{chemneq}).
These formulae can be efficiently calculated through a Bessel function expansion \cite{Fer50,share}.

Note that the series start diverging at $\lambda_i \rightarrow e^{m_i/T}$, and that for fluctuations the higher order terms in the expansion are always
more significant than for yields: for $\lambda_i  e^{m_i/T} = 1-\epsilon$
% then
\begin{equation}
\label{divergence}
\lim_{\epsilon\rightarrow 0} \ave{ N_i} \propto \epsilon^{-1},
\qquad
 \lim_{\epsilon\rightarrow 0} \ave{\Delta N_i^2} \propto \epsilon^{-2}.
\end{equation}
The parameter $\lambda_i$ corresponds to the particle fugacity, related to the chemical potential by $\lambda_i=e^{\mu_i/T}$.  Provided the law of mass action holds, it should be given by the product of charge fugacities (flavor, isospin etc.).  It is then convenient to parametrize it in terms of equilibrium fugacities $\lambda_i^{\mathrm{eq}}$ and phase space occupancies $\gamma_i$.   For a hadron with $q (\overline{q})$ light quarks, $s (\overline{s})$ strange quarks and isospin $I_3$ the fugacity is then
\begin{eqnarray}
\label{chemneq}
\lambda_i = \lambda_i^{\mathrm{eq}}
\gamma_q^{q+\overline{q}} \gamma_s^{s+\overline{s}}
\phantom{A},\phantom{A}
\lambda_i^{\mathrm{eq}}=\lambda_{q}^{q-\overline{q}} \lambda_{s}^{s -
\overline{s}}
\lambda_{I_3}^{I_3} 
\end{eqnarray}
If the system is in chemical equilibrium then detailed balance requires that $\gamma_q=\gamma_s=1$.   In an expanding system, however, the condition of chemical equilibrium 
no longer holds.  Kinetically, this occurs because collective expansion and cooling will make it impossible for endothermic and exothermic reactions, or for creation and destruction
reactions of a rare particle, to be balanced. Provided the system remains in local thermal equilibrium, $\lambda_i$ can still be used as a Lagrange multiplier for the
particle density, and the first and second cumulants can be calculated from the partition function the usual way \cite{jansbook,chemistrywebsite}.  However, in this case in general
$\gamma_q \ne 1,\gamma_s \ne 1$.
%%%%%%%%%%%%%%%%%%%%%%%%%%
\begin{figure}[tb]
\psfig{width=8.cm,clip=,figure=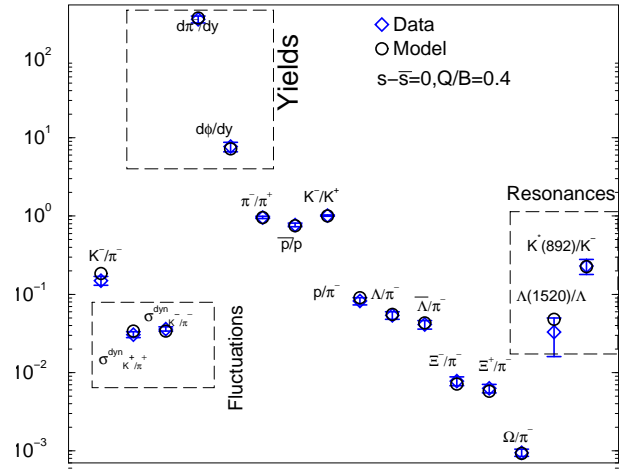}
\caption{\label{graph200}
Fit of preliminary 200 GeV data, including the $K^{\pm}/\pi^{\pm}$ fluctuations and the $K^* (892)$ and $\Lambda^* (1520)$ resonance}
\end{figure}
%%%%%%%%%%%%%%%%%%%% 

This picture becomes particularly appropriate if the expanding system undergoes a fast phase transition from a QGP to a hadron gas.  In this case, chemical non-equilibrium  \cite{sudden} and super-cooling \cite{csorgo} can arise due to entropy conservation:
By dropping the hadronization temperature to $\sim 140 MeV$ and oversaturating the hadronic phase space above equilibrium ($\gamma_q \sim 1.5,\gamma_s \sim 2$), it is possible to match the entropy of a hadron gas with that of a system of nearly massless partons \cite{sudden}.

Fits to experimental data  at both SPS \cite{observing}
and RHIC energies \cite{Rafelski:2003ju} indeed support these values of $\gamma_{q,s}$ when these parameters are fitted.  Moreover, best fit $\gamma_{q,s}>1$ arises for a critical energy \cite{gammaq_energy}  (corresponding to the energy of the $K/\pi$ ``horn'' \cite{horn}) and system size \cite{gammaq_size}, just as expected from the interpretation of $\gamma_q$ as a manifestation of a phase transition.  However, the fits performed in \cite{gammaq_energy} have not been able to rule out  equilibrium models (at SPS and RHIC the difference in statistical significance between equilibrium and non-equilibrium is $\sim 20 \%$).  

In the absence
of contrary evidence, equilibrium models are usually preferred for their 
smaller number of free parameters \cite{barannikova,bdm}.  Equilibrium freeze-out temperature varies between fits, ranging from 155 \cite{gammaq_energy} to 177 \cite{bdm} MeV.

Both scenarios are physically reasonable, can describe the data, and would be instrumental in our understanding of hadronic matter if proven correct.   In particular, the HBT puzzle suggests we lack understanding of the last stages of the fireball evolution.    Non-equilibrium is useful in this respect, since it affects both system volume \cite{bari,gammaq_energy} and emission time \cite{csorgo}.

The reason both equilibrium and non-equilibrium are compatible with data is that in a fit to yields the non-equilibrium phase space occupancies
$\gamma_s$ and $\gamma_q$ correlate with freeze-out temperature  \cite{gammaq_energy}, making a distinction between
 a $T=170$ MeV equilibrated freeze-out ($\gamma_q=1,\gamma_s\leq 1$) scenario and a supercooled scenario where $\gamma_{q,s}>1$ problematic.  

A related ambiguity is the difference between chemical freeze-out (where particle abundances are fixed) and thermal
freeze-out (where particles cease to interact).   Equilibrium models generally assume a long phase between these two points, which would alter considerably the multiplicity of directly detectable resonances.  In a Non-equilibrium supercooled freeze-out, on the other
hand, it is natural to assume that particle interaction after emission is negligible.   Once again, a reliable way to probe the extent of the
reinteraction would be instrumental for our understanding of how the fireball produced in heavy ion collisions breaks up.

We have recently shown \cite{prlfluct} that event-by-event fluctuations can be used to solve both of the dilemmas discussed above.
From the equations in the previous section, it is clear that the dependence of fluctuations of $T$ and $\gamma_q$ is different, allowing us to decouple these two variables.  A higher temperature tends to decrease fluctuations w.r.t. the Poisson value expected from Boltzmann statistics
($\sigma = 1$)
, since it introduces greater correlations due to an increased
contribution of resonances ($D_{12}$ in Eq. \ref{fluctratio}) .  Vice-versa, increasing $\gamma_q$ will rapidly increase fluctuations of quantities related to pions, due to the fact
that at $\gamma_q>1$ $\lambda_{\pi}$ rapidly approaches $e^{m_{\pi}/T}$, giving fluctuations an extra boost w.r.t. yields as per Eq. \ref{divergence}

In addition, comparing fluctuations to directly detected resonances probes the interval between chemical and thermal freeze-out.
Consider, for example, the $K^+/\pi^-$ fluctuation.
The top and the bottom terms in this ratio are linked by a large correlation term due to the $K^{*0}(892)$ decay.   
This correlation probes the $K^{*0}$ abundance at {\it chemical } freeze-out, since subsequent rescattering/regeneration does not alter the fact that the  $K^{*0}$ decay produced a $\pi^+$ and a $K^-$.
On the other hand, a direct measurement of the $K^*/K^-$ ratio through invariant mass reconstruction measures the $K^*$ abundance at {\em thermal} freeze-out, after all rescattering.
Hence, comparing the $K/\pi$ fluctuation   to the  $K^*/K^-$ ratio provides a gauge for effect of the hadronic reinteraction period on particle abundances.   In particular, a sudden freeze-out model should describe both with the same parameters.

The main experimental problem with fluctuation measurements is the vulnerability to effects resulting from limited detector acceptance.
This difficulty can be lessened, to some extent, by considering ``dynamical'' fluctuations, obtained by subtracting a ``static'' contribution which
should be purely Poisson in an ideal detector.   The exact nature of the subtraction varies from collaboration to collaboration, for example \cite{supriya} using
\begin{equation}
\sigma^{dyn}=\sqrt{\sigma^2-\sigma_{stat}^2}
\end{equation}
$\sigma_{stat}$, usually obtained through a Mixed event approach \cite{pruneau},
includes a baseline Poisson component, which for a ratio $N_1/N_2$ can be modeled as
\begin{equation}
\label{poissrat}
\sigma_{stat}^2 = \frac{1}{\ave{N_1}}+\frac{1}{\ave{N_2}}
\end{equation}
 as well as a contribution from detector efficiency and kinematic cuts.    Provided certain assumptions for the detector response function hold (see appendix A of \cite{pruneau}), subtracting
$\sigma_{stat}$ from $\sigma$ should yield a ``robust'' detector-independent observable.

When using measured $\sigma^{dyn}$s in fits to experimental data, the data-sample should include $\sigma^{dyn}$, particle yields (which are needed to determine the Poisson contribution to $\sigma_{stat}$ as per Eq. \ref{poissrat}) and particle ratios.
%%%%%%%%%%%%%%%%%%%%%%%%%%
\begin{figure}[tb]
\epsfig{width=8.cm,clip=,figure=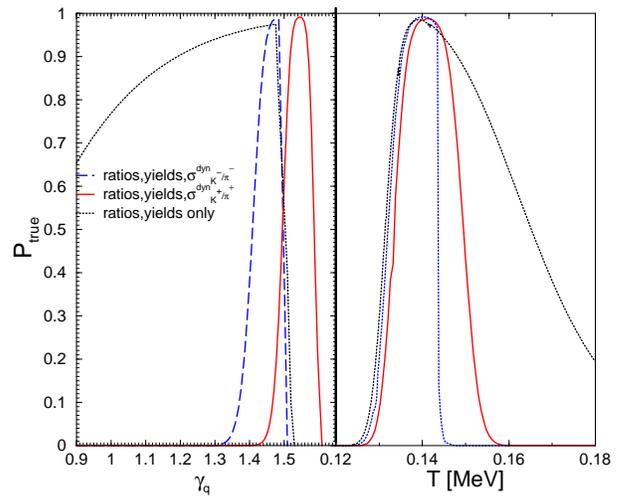}
\caption{\label{prof_200}
Temperature and $\gamma_q$ statistical significance ($P_{true}$) profiles for the fit in Fig. 1.  At each point in the abscissa a fit is performed varying all fit parameters except the one on the abscissa.  $P_{true}$ is calculated from the best fit parameters }
\end{figure}
%%%%%%%%%%%%%%%%%%%%%

We have performed a fit incorporating all ratios from \cite{barannikova}, with the exception of the $\Delta^{++}/p$, which the STAR collaboration has taken back, and the
$\overline{\Omega}/\Omega$, measured by STAR to be above unity (The latter measurement is a priori incompatible with statistical production \cite{liu}).  In addition, we have included the preliminary value
for the $K/\pi$ event-by-event fluctuation measured by STAR \cite{supriya}, as well as the published yield for $\phi$ \cite{starphi200} and $\pi^-$ \cite{starpi200}.

Our fit parameters include the normalization (hopefully related to the system ``volume'' at chemical freeze-out), temperature, $\lambda_{q,s,I_3}$ and $\gamma_{q,s}$.
We also require, by implementing them as data-points, strangeness, charge and baryon number conservation\\ ($
 \ave{s-\overline{s}}=0 \phantom{AA}\ave{Q}/\ave{B}=(\ave{Q}/\ave{B})_{Au}=0.4.)$.

The result is shown in Fig. \ref{graph200}.    The fit gives an adequate description of all particle yields, including the
resonance $K^*/K^-$ and $\Lambda(1520)/\Lambda$.   It can also, adequately describe the event-by-event fluctuations of of $K^+/\pi^+$ \textit{or} $K^-/\pi^-$, but not the difference
between them.   Describing the difference within the SHM is impossible without introducing
a sizable isospin fugacity $\lambda_{I3} \sim 0.96$ which is in turn excluded by ratios such as $\pi^+/\pi^-$ and $Q/B$.
We wait for published results, and systematic error computation before studying this discrepancy further.

Fig. \ref{prof_200} shows that the correlation between $T$ and $\gamma_q$ disappears when fluctuations are taken into account, and that the fit tightly constrains $\gamma_q$ well above
the equilibrium value, forcing temperature to be about 140 MeV, in good agreement of the prediction of the supercooled hadronization scenario.  In accordance with this scenario, our model
can describe both the $K/\pi$ fluctuation and the directly observed $K^*/K^-$ value with the same statistical parameters, as is expected for a system where
hadronic reinteraction (the phase between chemical freeze-out and thermal freeze-out) is negligible.

\begin{figure}[tb]
\epsfig{height=7.cm,clip=,figure=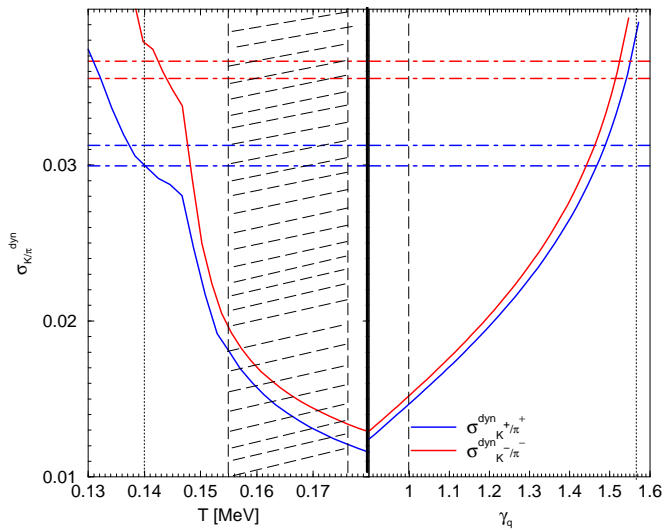}
\caption{\label{datkpi_200}
Sensitivity of the $K/\pi$ fluctuation $(\sigma_{K/\pi}^{dyn})$ to $\gamma_q$ and temperature  for the fit in Fig. 1.  At each point in the abscissa, a fit is performed (only yield data-points used) varying all fit parameters except the one on the abscissa. $\sigma_{K/\pi}^{dyn}$ is calculated using the best fit parameters.
dot-dashed lines refer to the experimental limits \cite{supriya}  for the $\sigma_{K^+/\pi^+}$ (blue) and $\sigma_{K^-/\pi^-}$ (red).
Dotted vertical lines give published non-equilibrium best fit parameters \cite{gammaq_energy}, while the long-dashed line and shaded area gives published equilibrium best fit \cite{barannikova,bdm,gammaq_energy}.  }
\end{figure}

Fig. \ref{datkpi_200} confirms that fluctuations are indeed responsible for the tight constraining of $T$ and $\gamma_q$.  The best fit at light quark equilibrium misses $\sigma_{K/\pi}^{dyn}$
by several standard deviations.     Introducing exact conservation for strangeness within the observed window would increase the discrepancy by reducing $\sigma_{K}$ \cite{nogc2}.   It is only through $\gamma_q>1$ that $\sigma_{K/\pi}^{dyn}$ increases until it becomes compatible with the experimental value. 

%%%%%%%%%%%%%%%%%%%%%%%%%%
%%%%%%%%%%%%%%%%%%%%%

In conclusion, we have used preliminary experimental data to show that, at 200 GeV, the equilibrium model is unable to describe both yields and fluctuations within the same statistical parameters.
The non-equilibrium model, in contrast, succeeds in describing almost all of the yields and fluctuations measured so far at RHIC, with the parameters expected from a scenario where
non-equilibrium arises through a phase transition from a high entropy state, with super-cooling and oversaturation of phase space.   We await more published data to determine weather the non-equilibrium
model is really capable of accounting for both yields and fluctuations in all light and strange hadrons produced in heavy ion collisions.

Work supported in part by grants from
the U.S. Department of Energy  (J.R. by DE-FG02-04ER41318)
the Natural Sciences and Engineering research
council of Canada, the Fonds Nature et Technologies of Quebec.
G. T. thanks the Tomlinson foundation  for support, and the QM2005 
organizers for financial support during the conference.
S.J.~thanks RIKEN BNL Center
and U.S. Department of Energy [DE-AC02-98CH10886] for
providing facilities essential for the completion of this work.

\end{document}